\documentclass[twocolumn]{revtex4}
\usepackage{epsfig}
\usepackage{graphicx}
\usepackage{amsfonts,amssymb}
\usepackage{amsmath}
\usepackage{color}
\usepackage {times}
\usepackage{lipsum}
\usepackage{bm}% bold math
\usepackage{blindtext}
\usepackage[title]{appendix} 
\usepackage[utf8]{inputenc}
\usepackage[T1]{fontenc}
\DeclareUnicodeCharacter{0308}{}
\definecolor{refcolor}{rgb}{1.0,0.0,0.0}
\newcommand{\be}{\begin{equation}}
\newcommand{\ee}{\end{equation}}   
\newcommand{\bea}{\begin{eqnarray}}
\newcommand{\eea}{\end{eqnarray}}
\newcommand{\ba}{\begin{array}}
\newcommand{\ea}{\end{array}}

\renewcommand{\k}{{\bf k}}

\begin{document}

\title{Antiferromagnetically ordered Dirac semimetal in Hubbard model with spin-orbit coupling}

\author{Garima Goyal and Dheeraj Kumar Singh}

\address{Department of Physics and Materials Science, Thapar Institute of Engineering and Technology, Patiala-147004, Punjab, India}     
\date{\today}

\begin{abstract}
We examine the possible existence of Dirac semimetal with magnetic order in a two-dimensional system with a nonsymmorphic symmetry by using the Hartree-Fock mean-field theory within the Hubbard model. We locate the region in the second-neighbor spin-orbit coupling vs Hubbard interaction phase diagram, where such a state is stabilized. The edge states for the ribbons along two orthogonal directions concerning the orientation of in-plane magnetic moments are obtained. Finally, the effect of the in-plane magnetic field, which results in the stabilization of the Weyl semimetallic state, and the nature of the edge states corresponding to the Weyl semimetallic state for ribbon geometries are also explored.
\end{abstract}

\maketitle
\section{Introduction}
Significant progress, accomplished in the direction of topological band theories of insulators~\cite{kane, hasan, nakahara, fu, bernevig, haldane} and superconductors~\cite{qi, bernevig1, schnyder, kitaev, qi1, bansil} in the last decade, has inspired parallelly the study of gapless systems with nontrivial topology~\cite{deb, kane1, roy, xu, xia, wu}. The crucial band feature of the gapless topological systems, \textit{i.e.}, semimetals is the symmetry-protected band crossing at the Fermi energy~\cite{yang, weng, weng1, fang, yan, armitage, gao, young}. There exists a non-zero topological charge associated with the flux of a curvature function such as the Berry curvature along a closed loop in the Brillouin zone enclosing the band-crossing point~\cite{berry, burkov}. These semimetals  exhibit a variety of electronic and transport properties including the edge states~\cite{armitage, burkov1, li1}, anomalous Hall effect~\cite{burkov}, large thermopower~\cite{huang} etc., which make them important for a variety of technological applications~\cite{liu, lundgren}.  

Graphene, with a two-sublattice and  honeycomb lattice crystal structure, is one of the systems studied extensively for its topological semimetallic state, which is protected by time-reversal $(\mathcal{T})$ and parity ($\mathcal{P}$) invariance in the absence of spin-orbit coupling~\cite{castro, novoselov, zhou}. The quasiparticle dispersion, in the vicinity of band crossings at the Fermi level, is linear in all the momentum directions of the Brillouin zone. A linear energy dispersion near the band crossing points or the Dirac points (DP) describes a pseudo-relativistic massless Dirac fermion~\cite{wang1, young}. However, a small spin-orbit coupling (SOC) can destroy the Dirac cone, which results in an insulating state with a topological origin~\cite{kane, kane1}. Intense effort is underway to search for 2D material systems besides graphene, which can support the symmetry-protected Dirac semimetal (DSM), because of the tunability of the physical properties of two-dimensional systems.

Recent studies~\cite{young1, wang, matneeva} suggest that the symmetry-protected two-dimensional topological semimetallic state (TSM) may also exist in the presence of SOC unlike in the case of  graphene. However, such a  TSM state requires the presence of non-symmorphic crystalline symmetries. The two successive symmetry operations, which may be associated with non-symmorphic symmetry, belong to a point group such as rotation about an axis or reflection in a plane together with a fractional Bravais lattice translation~\cite{weider, zhao1, bradley}. 

 DPs can also be stabilized in three-dimensional systems~\cite {wang, wang2}. 3D Dirac semimetal (DSM) was proposed earlier in $\beta$-cristobalite BiO$_2$~\cite{young} and experimental signatures were obtained later in materials such as Na$_3$Bi~\cite{liu1}, Cd$_3$As$_2$~\cite{liu2,borisenko} etc. There are two routes via which 3D DSMs can be obtained~\cite{yang}. First, such a state can appear as an intermediate state, when a system is driven across a transition from the trivial to a topological insulating state in the presence of band inversion. Here, the stability of DPs requires inversion symmetry in addition to the time-reversal symmetry and the removal of any of these two symmetries leads to a semimetal with DPs being split into the Weyl points with non-zero Chern number~\cite{murakami, wan, xu1, burkov2, su, lv}. The other route is through the touching of conduction and valence bands at discrete points in the Brillouin zone, where the DPs are protected by the non-symmorphic symmetries of the crystal-space group. 3D DSM state may also exist in the absence of $\mathcal{T}$ or $\mathcal{P}$ with candidate materials suggested to be CuMnAs and CuMnP with an antiferromagnetic order (AFM)~\cite{tang}. Protection of DPs in 2D DSM state with AFM order, in the presence of glide mirror symmetry consisting of reflection about a plane and half translation, was also explored, when both $\mathcal{T}$ and $\mathcal{P}$ are broken~\cite{wang}.     

The stability of the 2D DSM or Weyl semimetallic (WSM) state with AFM order may require a rather stringent condition on the orientation of magnetic moments to be fulfilled. Such a state is possible only when the magnetic moments are aligned along the plane of the two-dimensional system in a way that they are directed along the line joining nearest neighbor atoms on a given sublattice ~\cite{wang}. However, it is yet to be seen whether this type of state can be realized in a one-orbital tight-binding model with  Hubbard interaction. In this paper, we examine the possible existence of DSM state with AFM order within the Hubbard model in a setting, where both $\mathcal{T}$ and $\mathcal{P}$ are broken, while their combination $\mathcal{T}\mathcal{P}$ remains intact. When the magnetic moments lie in the plane, a phase diagram is obtained in the SOC vs interaction parameter space and the corresponding edge states are obtained. The consequence of magnetic field on the DPs and the modified edge states is studied.     

This paper is organized as follows. Section II describes the one-orbital Hubbard model with second-neighbor SOC in the sublattice basis. With mean-field decoupled Hubbard interaction, the Hamiltonian for magnetic moments pointing in arbitrary directions is presented. Section III discusses the results including a phase diagram in order to identify the region, where DSM with AFM order is stabilized, especially when the magnetic moments are oriented along the crystal plane. In addition, the edge states for ribbon geometry when the ribbon is oriented along a direction in which the magnetic moments are aligned and a direction perpendicular to it are studied. We also discuss the consequence of magnetic field and possible stabilization of WSM state. Finally, we present conclusions in section IV.
\section{Model and Method} 
\begin{figure} 
\centering
\includegraphics[width=0.80\linewidth]{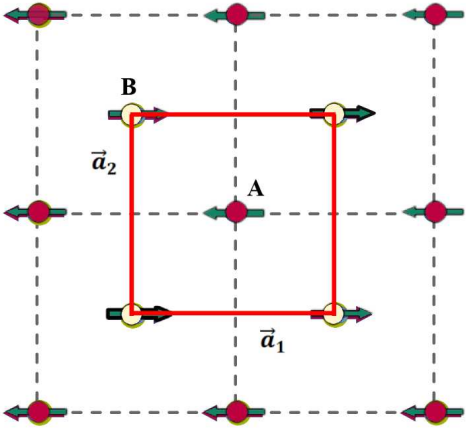}
\caption{AFM lattice arrangement with 2 atoms per unit cell (shown in red square) having sublattices A and B with magnetic moments aligned along $x$-direction whose lattice vectors are $\Vec{a}_1 = (1,0,0)$ and $\Vec{a}_2 = (0,1,0)$, respectively. The A(B) atoms are displaced along $+z(-z)$  directions by a tiny fraction of the lattice vector to make the crystal non-symmorphic. The red (yellow) circles show the atoms lying above (below) the plane.}
\label{1}
\end{figure}
In order to study the antiferromagnetic DSM state, we consider a tight-binding Hamiltonian in two-dimension with spin-orbit coupling~\cite{kane} and Hubbard interaction, which is given by 
\begin{eqnarray}
    H &=&  t\sum_{\langle {\bf i, j}\rangle , \sigma}  (a_{{\bf i} \sigma}^{\dagger} a^{}_{{\bf j}\sigma} + h.c. )+  \sum_{\bf i}  U n_{{\bf i} \uparrow} n_{{\bf i} \downarrow} \nonumber\\ &+& i \lambda \sum_{{\bf i}} \sum_{\sigma \sigma^{\prime}} (\nu_{{\bf i},{\bf i+ {\delta^{\prime}}}} a_{{\bf i} \sigma}^{\dagger} a_{{\bf i+\delta^{\prime}} \sigma^{\prime}}^{} \sigma_{\sigma \sigma^{\prime}} + h.c. )
\end{eqnarray}
The first term represents the delocalization energy gain arising because of hopping of an electron with the spin $\sigma$ from the site ${\bf i} $ to a nearest-neighbor site ${\bf j}$, where $a_{{\bf i } \sigma}^{\dagger} (a^{}_{{\bf i} \sigma})$ is the electron creation (destruction) operator. $t$ is the hopping parameter, which is set to be the unit of energy throughout unless mentioned otherwise. Here, only the nearest-neighbor hopping is considered as the second-neighbor hopping will break the particle-hole symmetry, which leads to two DPs appearing at different energies. 
The second term denotes the onsite Coulomb repulsion between the electrons of opposite spins, where  $n_{{\bf i} \uparrow} = a^{\dagger}_{{\bf i} \uparrow} a^{}_{{\bf i} \uparrow}$. The third term denotes the next-nearest neighbor SOC whereas the nearest-neighbor SOC is ignored. ${\bf \delta}^{\prime} = +\hat{x}$ or $+\hat{y}$. $\lambda$ is the SOC parameter and ${\nu}_{{\bf i}, {\bf i + {\delta}^{\prime}}}$ = -${\nu}_{ {\bf i +{\delta}^{\prime}},{\bf i}}$  = $\pm 1$ depending on the orientation of two next-nearest neighboring bonds. 

We consider a square lattice as shown in Fig.~\ref{1} with a single orbital per site.  The unit cell has two atoms placed in an AFM arrangement with the lattice vectors denoted by $\Vec{a_1} = (1,0,0)$ and $\Vec{a_2} = (0,1,0)$, respectively. In order to make the lattice non-symmorphic, the atoms in the unit cell namely A and B are displaced along $+z$ and $-z$ directions, respectively, by a tiny but equal fraction of the lattice vector. This atomic arrangement allows next-nearest-neighbor SOC~\cite{kane1}. 

The meanfield decoupling of the Hubbard term of the Hamiltonian given by Eq. 1 yields  
\begin{equation}
{H}_{im} 
=  - \frac{U}{2} \sum_{i \sigma} \Psi^{\dagger}_{i}(\bm{ \sigma}\cdot{\bf m}_{i}) \Psi_{i}
+ \frac{U}{4}\sum_{i}{\bf m}_{i}^{2}
\vspace{-3mm}
\end{equation}
{where $\Psi^{\dagger}_i = (a^{\dagger}_{i \uparrow}, a^{\dagger}_{i \downarrow} )$}. The $j$-th component of magnetic moment at the site $i$ is ${ m}^{j}_i = \frac{1}{2} \langle \Psi^{\dagger}_{i}{\sigma}^{j} \Psi_{i} \rangle$ whereas ${\bf m}_{i}$ is the magnetic moment at site $i$. ${\sigma}^{j}$ is $j$-th component of Pauli matrices. When incorporated into the sublattice structure of the system with AFM order under consideration, the meanfield decoupled part with broken $\mathcal{P}$ and $\mathcal{T}$ becomes
\begin{equation}
    H_{mf} = -\sum_{{\bf i} \alpha \beta}  \tau a_{{\bf i} \alpha}^{\dagger} (\bm{ \sigma} \cdot \bm{\Delta}_i)_{\alpha \beta}  a^{}_{{\bf i} \beta} ,
\end{equation}
where $\bm {\Delta}_i$ is the exchange field given by 
\begin{equation}
2\bm{\Delta}_i = U(m_{ix}\hat{x}+ m_{iy} \hat{y} + m_{iz} \hat{z}). 
\end{equation}
$m_{ix}$, $m_{iy}$, and $m_{iz}$ represent the components of magnetic moments, which are obtained self consistently. $\hat{x}$, $\hat{y}$, and $\hat{z}$ are the unit vectors. The last term of $H_{im}$ does not appear in $H_{mf}$ as it is a scalar field and does not play any role in the self-consistent calculation. $\tau = 1 $ on $A$ sublattice and -1 on $B$ sublattice. Thus, after Fourier transforming Eq. 1 and combining it with Eq. 3, the Hamiltonian in the ${\bf k}$-space can be written in the composite sublattice and spin basis as
\begin{eqnarray}
    \mathcal{H}({\bf k}) &=& 4t \tau_1 \cos \frac{k_x}{2} \cos \frac{k_y}{2} + (\Delta_x - 2 \lambda \sin k_y) \sigma_1 \otimes \tau_3 \nonumber\\ & +& ( \Delta_y + 2 \lambda \sin k_x) \sigma_2 \otimes \tau_3 +
    \Delta_z \sigma_3 \otimes \tau_3,
   \label{eqn.4} 
\end{eqnarray}
where $\sigma$ and $\tau$ are Pauli matrices in the spin and sublattice spaces. The two-fold degenerate eigenvalues of the Hamiltonian can be readily shown to be 
\begin{equation}
E_{\bf k} = \pm
\sqrt{ \varepsilon^2_{\bf k} + \Delta^{\prime 2}_{x \bf k} +\Delta^{\prime 2}_{y\bf k} + \Delta^2_z},
\end{equation}
where $\varepsilon_{\bf k} = 4t \cos \frac{k_x}{2} \cos \frac{k_y}{2}$, $\Delta^{\prime }_{x \bf k} = \Delta_x - 2 \lambda \sin k_y$,  $\Delta^{\prime }_{y\bf k} = \Delta_y + 2 \lambda \sin k_x$.
The eigenvector $(\phi^{A}_{{\bf k}\uparrow}$, $\phi^{B}_{{\bf k} \uparrow}$, $\phi^{A}_{{\bf k} \downarrow}$, $\phi^{B}_{{\bf k} \downarrow})^{T}$ of the Hamiltonian $\mathcal{H}({\bf k})$ is used to obtain the magnetic moments in a self-consistent manner, where one electron per site is considered throughout. The components of magnetic moments at the sublattice $A$ is 
\begin{eqnarray} 
m_z &=& n^A_{ \uparrow} - n^A_{ \downarrow} \nonumber\\
&=& \sum_{{\bf k},l}(\phi^{A*}_{ {\bf k}\uparrow} \phi^{A}_{{\bf k}\uparrow} \Theta(E_f - E_{\k, \uparrow,l})- \phi^{A*}_{ {\bf k}\downarrow} \phi^{A}_{{\bf k}\downarrow}\Theta(E_f - E_{\k, \downarrow, l}))  \nonumber\\
m_x &=& \sum_{{\bf k},l}(\phi^{A*}_{{\bf k}\uparrow} \phi^{A}_{{\bf k}\downarrow} + \phi^A_{{\bf k}\uparrow} \phi^{A*}_{{\bf k}\downarrow}) \Theta(E_f - E_{\k, \uparrow,l})\Theta(E_f - E_{\k, \downarrow,l}) \nonumber\\
m_y &=& \sum_{{\bf k},l}(-i\phi^{A*}_{ {\bf k}\uparrow} \phi^{A}_{{\bf k}\downarrow} + i \phi^{A*}_{ {\bf k}\uparrow} \phi^{A}_{{\bf k}\downarrow}) \Theta(E_f - E_{\k, \uparrow,l}) \nonumber\\ &\times&\Theta(E_f - E_{\k, \downarrow,l}), \nonumber\\
\end{eqnarray}
where $l$ is the band index.

\section{Results}
The energy dispersion given by Eq. 6 is gapped whenever the magnetic moments have a component along $z$ direction or along a direction in the $x$-$y$ plane other than $x$ or $y$ axes. However, there exists a two-fold degenerate band crossing at $E_k = 0$ provided that the magnetic moments are oriented either along $x$ or $y$ direction. In these cases, the energy dispersions are linear in the vicinity of the band crossing which indicates that these crossing points are DPs. These DPs are protected by glide mirror symmetry $\{M_{\hat{x}}|\frac{1}{2}0\}$ and $\{M_{\hat{y}}|0\frac{1}{2}\}$ depending on whether the magnetic moments are oriented along $x$ and $y$, respectively~\cite{wang}. 
\begin{figure} 
\centering
\includegraphics[width=0.90\linewidth]{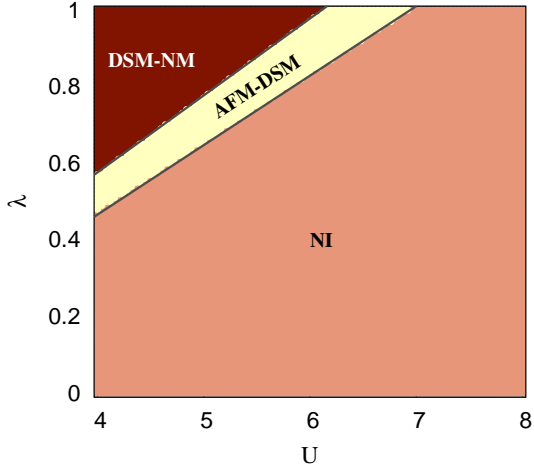}
\caption{Phase diagram in the  $\lambda-U$ space in the range $4  \le U \le 8$ and $0 \le \lambda \le 1.0$ at half-filling ($n = 1.0$). DSM state with AFM order is obtained only for a very narrow range of $\lambda$ which separates the DSM state without magnetic order (DSM-NM) and the AFM insulating state (AFM-I).}
\label{2}
\end{figure}

Fig.~\ref{2} shows the self-consistently obtained phase diagram in the $\lambda$-$U$ parameter space when the magnetic moments are aligned along the $x$ direction. Three different phases are obtained; DSM without magnetic order, DSM with AFM order, and a normal insulator with AFM order. It may be noted that magnetic-order parameter vanishes in the DSM-NM state found in the small $U$ but large $\lambda$ region. Thus, there is no order parameter associated with this phase and both inversion and time-reversal symmetry are intact and DPs are protected by the nonsymmorphic symmetries. However, the magnetic-order paramater is non zero and time-reversal symmetry is broken in the DSM state with AFM order (AFM-DSM) as well as in the normal insulator with AFM order (AFM-I). The latter two are differentiated only by condition based on the relative strength of magnetic-moment dependent exchange coupling and SOC, which is discussed later in this paragraph. Unlike in the symmorphic system, there is no further reduction in the translational symmetry originating from the sublattice structure associated with AFM ordering of magnetic moments. This is because the atoms in the sublattices $A$ and $B$ are displaced by a tiny fraction of lattice vector in a direction perpendicular to the plane of two-dimensional system in the original model itself (Fig. 1). The DSM state with AFM order is obtained only for a very narrow window of SOC centered around $\lambda \sim$ 0.8, which is sandwiched in between the DSM state without magnetic ordering and the AFM insulating state (AFM-I). The DPs occur at the momenta $\textbf{k} = (\pi, k_{y0})$ and $(\pi, \pi -k_{y0})$ with $\sin k_{y0} = \Delta_x/2 \lambda$~\cite{wang}. \textcolor{blue}Therefore, whenever the self-consistently obtained exchange field satisfies the condition $0 < \Delta_x \le 2 \lambda$, the AFM-DSM state is obtained, $\Delta_x = 0$ corresponds to DSM-NM state, and $\Delta_x \ge 2 \lambda$ is satisfied in the AFM-I state. 

\begin{figure} 
\centering
\includegraphics[width=0.97\linewidth]{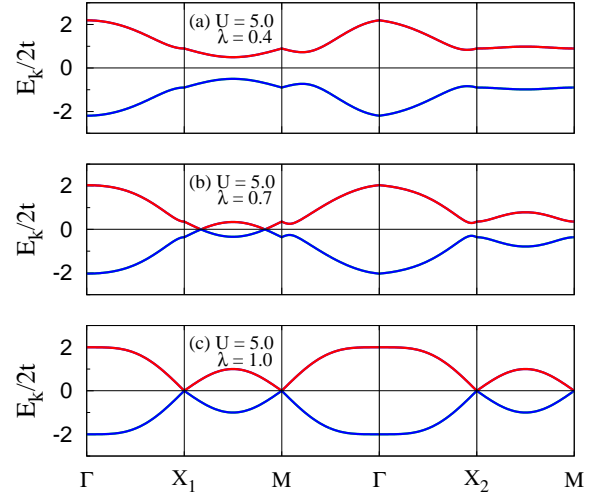}
\caption{For $U = 5$, electronic dispersions are plotted when (a) $\lambda = 0.4$, (b) $\lambda = 0.7$ and (c) $\lambda = 1.0$. For these three different values, normal insulator, DSM state with antiferromagnetic order and without any magnetic order are obtained, respectively. In each case, the DPs are protected by the non-symmorphic symmetries.}
\label{3}
\end{figure}

Fig.~\ref{3} shows the electronic dispersion in the three phases for a fixed Coulomb interaction parameter $U = 5.0$ and different values of SOC parameter $\lambda$. In Fig.~\ref{3}(a), the bulk dispersion is obtained for $\lambda = 0.4$ corresponding to the normal insulating state with AFM order indicated by a gap opening at the Fermi level. The gap opening disappears when SOC is increased and two DPs appear along X$_1$-M but they are located away from the high-symmetry points X$_1$ and M (Fig.~\ref{3}(b)). The energy dispersion in the vicinity of DPs can be obtained from Eq. (6), which is 
\begin{equation}
E_k = \pm 2\sqrt{(\lambda^2 + t^2 \cos^2 k_{y0}/2 )q_x^2 +(\lambda^2 \cos^2 k_{y0}) q^2_y}. 
\end{equation}
These DPs are protected by glide mirror plane symmetry $\{M_{\hat{x}}|\frac{1}{2}0\}$ while the dispersion in their vicinity is independent of magnetic moment. For $\lambda = 1.0$ (Fig.~\ref{3}(c)), the magnetic moment vanishes, therefore, three DPs are obtained at the time-reversal invariant momenta, \textit{i.e.}, X$_1(\pi,0)$, M($\pi, \pi)$ and X$_2(0, \pi)$ of the Brillouin-zone boundary. These DPs are protected by either of the screw axes $\{C_{2\hat{x}}|\frac{1}{2}0\}$ and $\{C_{2\hat{y}}|0\frac{1}{2}\}$ or by glide mirror plane symmetry $\{M_{\hat{z}}|\frac{1}{2}\frac{1}{2}\}$ as the system has now both $\mathcal{T}$ and $\mathcal{P}$ symmetry intact in the absence of any magnetic moment~\cite{kane1}. 
\begin{figure}[hb]
    \centering
    \includegraphics[width=1.05\linewidth]{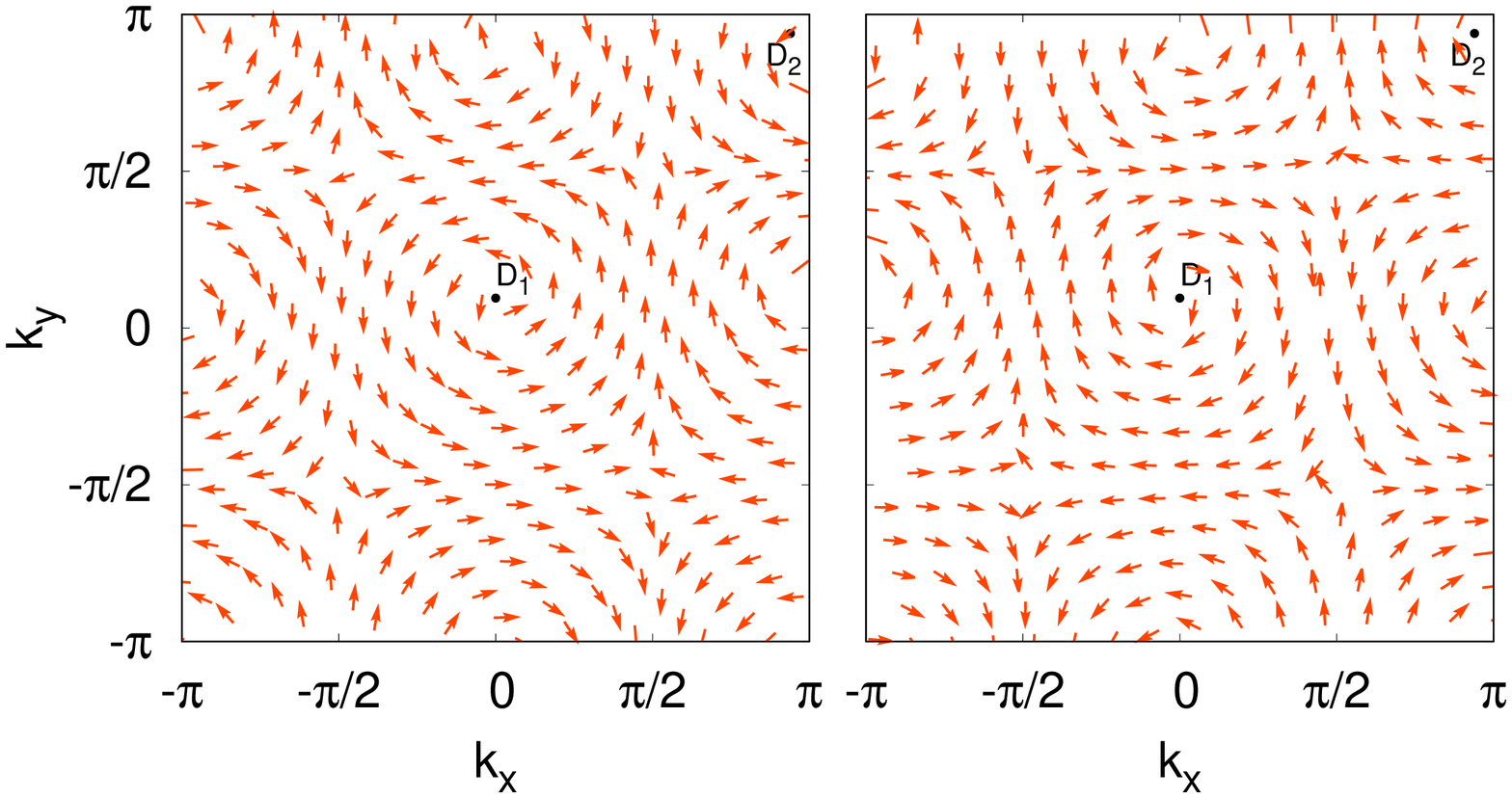}
    \caption{Berry connection plotted for the two individual bands in the whole Brillouin for the AFM-DSM state. The two DPs shown in Fig. 3(b) are denoted by $D_1$ and $D_2$. The Berry connection for the different bands has opposite signs as indicated by the clockwise and counterclockwise rotations in the two plots.}
    \label{4}
\end{figure}
The topological charge \textit{i.e.} the Chern number associated with the DPs should vanish as the degenerate bands have opposite chirality. The Chern number can be calculated with the help of the Berry flux, which is same as the line integral of the Berry connection along the boundary of a cross-section $\mathcal{S}$ in the ${\bf k}$ space, which includes one of the DPs. Thus, the Berry flux is given by
\begin{equation}
    \phi_n = \int_{\mathcal{S}} {\bf B}_n ({\bf k}) \cdot d {\bf S}
\end{equation}
where $n$ is the band number.
The berry curvature can be written in terms of Berry connection ${\bf A}_n ({\bf k})$ as 
\begin{equation}
    {\bf B}_n ({\bf k}) = {\bf \nabla} \times {\bf A}_n ({\bf k})
\end{equation}
or equivalently as a line integral due to Stoke's theorem. The Berry connection for the $n^{th}$ band is
\begin{equation}
    {\bf A}_n ({\bf k}) = i \langle \psi_n | {\bf \nabla_{k}} | \psi_n \rangle .
\end{equation}
Fig.~\ref{4}(a) and (b) show the Berry connection plotted for the AFM-DSM state in the entire Brillouin zone. The rotation of the Berry connection for the two degenerate bands around each of DPs are opposite to each other whereas the chirality of one of the bands is -1 while for the other it is 1. Thus, the Chern number vanishes, $\sum_n B_n ({\bf k}) = 0$, $\phi = \sum_n \phi_n = 0$ for each DPs.

The edge states, in the topological insulators, are pairs of states with opposite spins propagating in directions opposite to each other. The dispersion of the edge state crosses the Fermi level and appears as a bridge between the bands corresponding to valence and conduction electrons. They are also supported in systems such as Dirac and Weyl semimetals. There are numerous examples including the localized flat edge states in the quasi-one-dimensional graphene ribbons of zigzag shape~\cite{nakada, fujita, wakabayashi, wakabayashi1}. While the edge states in graphene are attributed to the Dirac cones, these states may even be found in their absence when degeneracy occurs at high symmetry points~\cite{fertig, lau, lado}. Recent studies have explored the edge states in the Weyl semimetals without any magnetic order when the time-reversal symmetry is broken in the nonsymmorphic system~\cite{matneeva}. 

Here, we examine the edge-state dispersion in the Dirac semimetallic state with AFM order in quasi-one dimensional system. It may be noted that the nature of the edge state may depend on how the in-plane magnetic moments are oriented with respect to the ribbon length, as the four-fold rotation symmetry is broken. First, we consider a ribbon of width $W$ lying along the $y$ direction so that $k_y$ becomes a good quantum number. The ribbon Hamiltonian $H_{Rby}$ with dimension $2W \times 2W$, where $W$ is the number of atomic chains in the ribbon, is given by

\begin{equation}
    H_{Rby} ({\bf k}) = 
    \begin{pmatrix}
        H_{1+}  &  H_2  &  H_3  &  O  &  \cdots \\
        H^{\dagger}_2  &  H_{1-}  &  H_2  &  H_3  &  \cdots \\
        H^{\dagger}_{3}  &  H^{\dagger}_2  &  H_{1+}  &  H_2  & \cdots \\
        O  &  H^{\dagger}_{3}  &  H^{\dagger}_2  &  H_{1-}  &  \cdots \\
        \vdots  &  \vdots  &  \vdots  &  \vdots  &  \ddots \\
    \end{pmatrix},
\end{equation}

where 

\begin{equation*}
    H_{1 \pm} = 
    \begin{pmatrix}
        0  &  \pm(-2 \lambda \sin k_y + \Delta_x)  \\
        \pm(-2 \lambda \sin k_y + \Delta_x)  &  0 \\
    \end{pmatrix},
\end{equation*}

\begin{equation*}
    H_2 = 
    \begin{pmatrix}
        2t \cos (k_y/2)  &  0  \\
        0  &  2t \cos (k_y/2)  \\
    \end{pmatrix}
\end{equation*}
and
\begin{equation*}
    H_3 = 
    \begin{pmatrix}
        0  &  -\lambda  \\
        \lambda  &  0  \\
    \end{pmatrix}.
\end{equation*}
$H_{1 \pm}$ is the element in a matrix form of Hamiltonian corresponding to a single chain while $H_{2 }$ and $H_{3 }$ matrices connect a chain to the nearest and the next-nearest neighbor chains.

\begin{figure}
\centering
\includegraphics[width=8.5cm,scale=1.0]{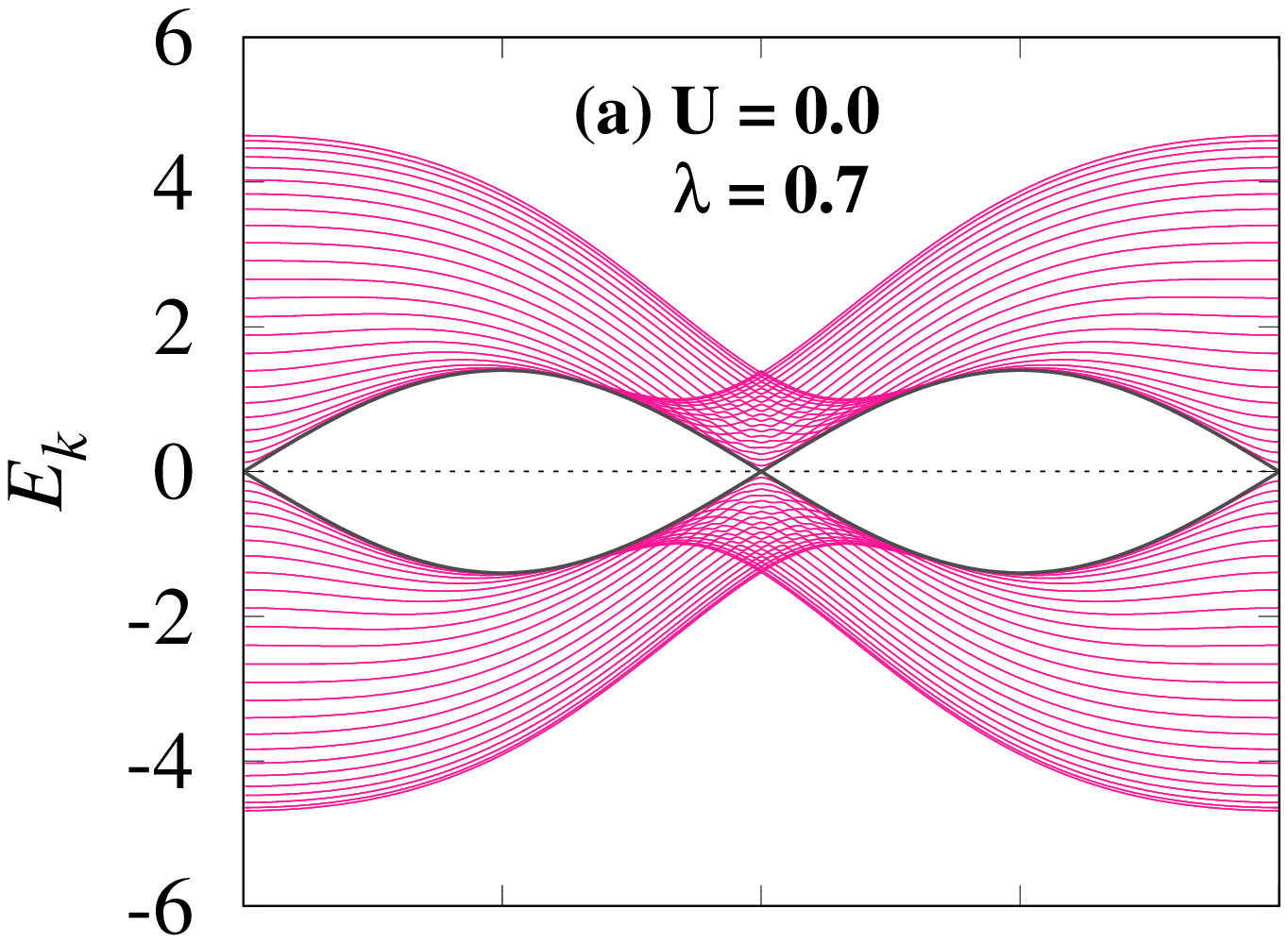}\\
\vspace*{-0.7cm}
\includegraphics[width=8.5cm,scale=1.0]{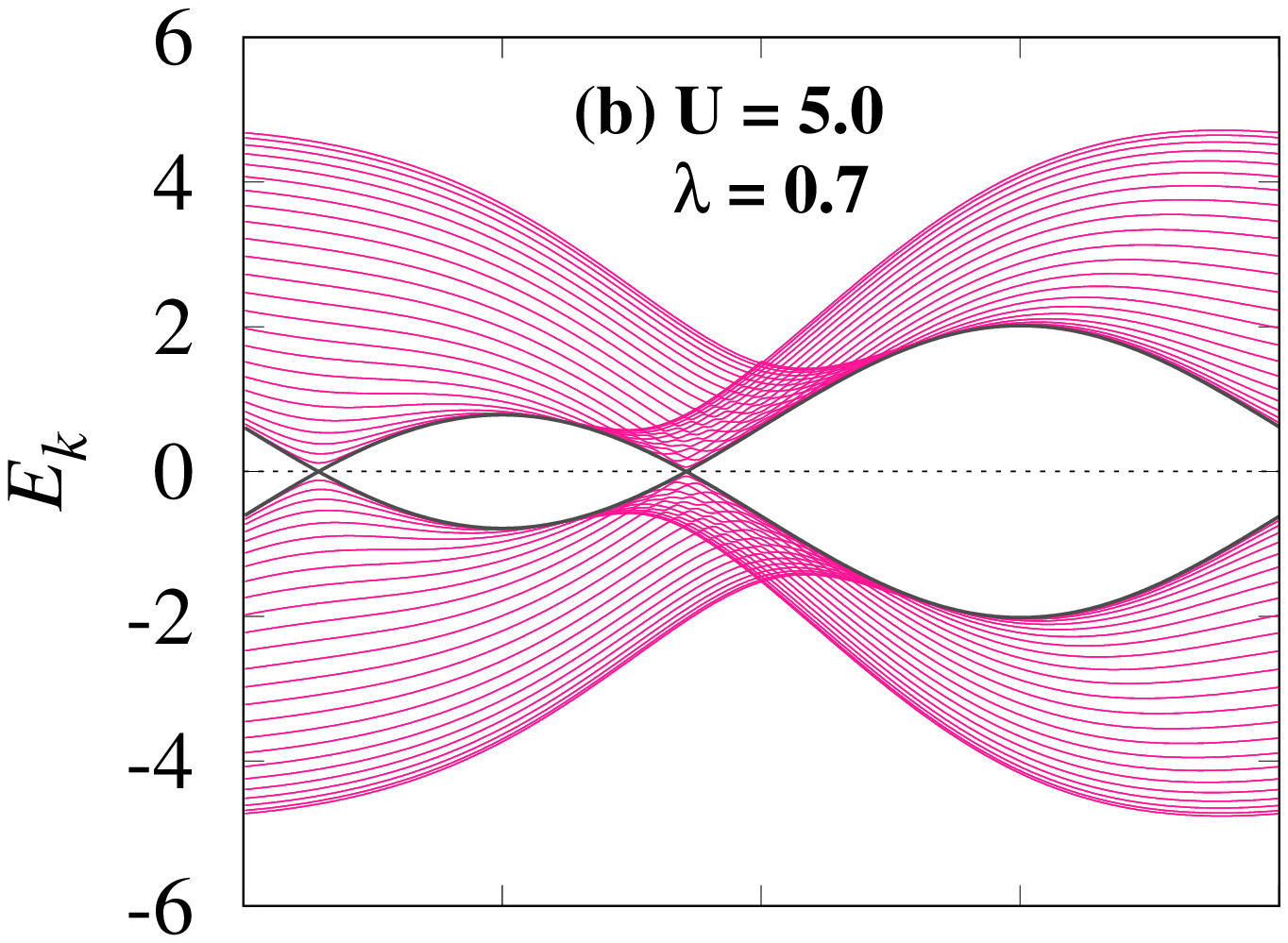} \\
\vspace*{-0.7cm}
\includegraphics[width=8.5cm,scale=1.0]{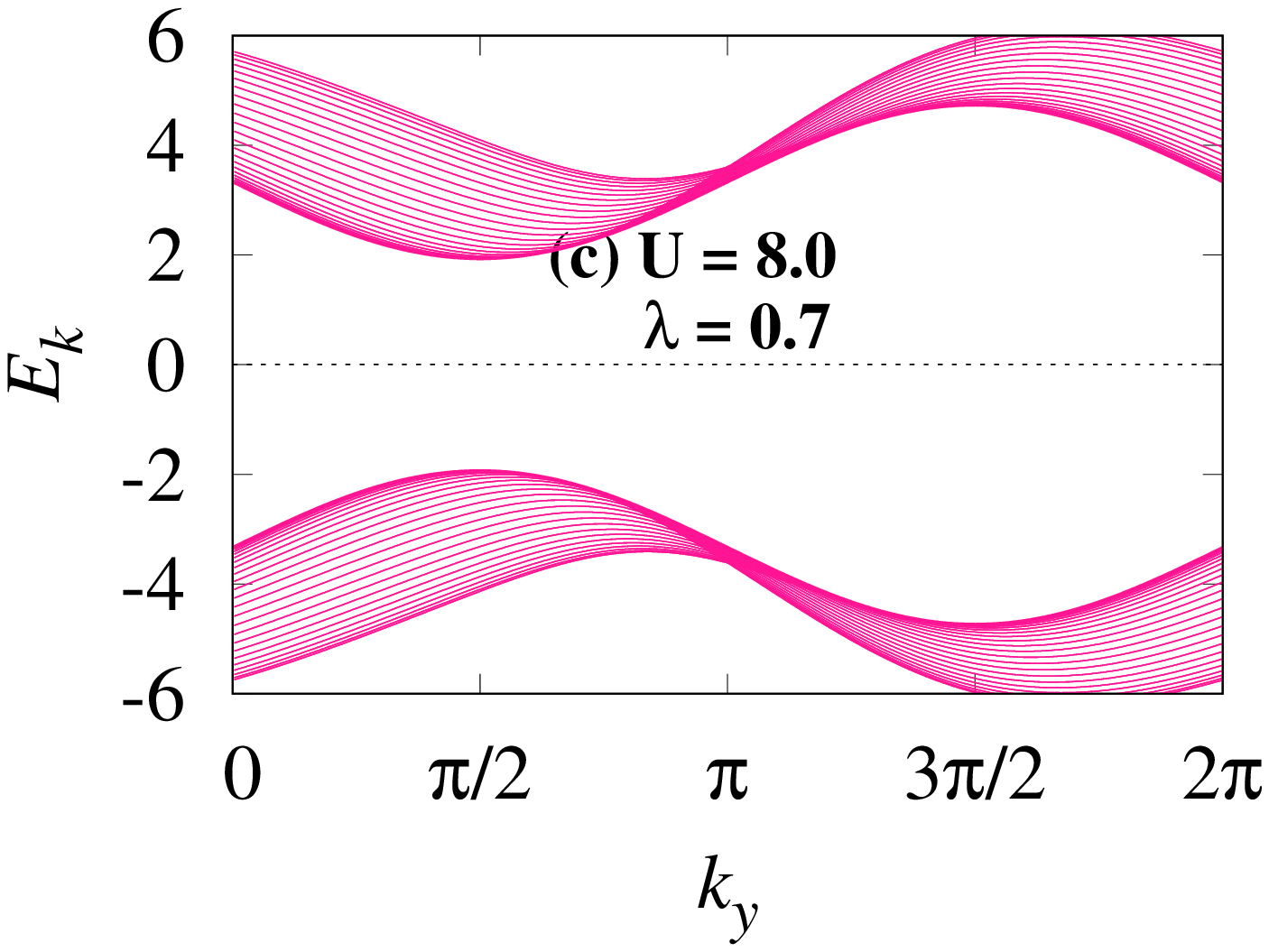} \\
\vspace*{-0.5cm}
\caption{Edge-state dispersions and bulk bands for a ribbon of width $W = 50$ lying along $y$ direction and projected onto one dimensional Brillouin zone for $\lambda=0.7$ and (a) $U=0.0$, (b) $5.0$ and (c) $8.0$. The edge-state dispersion crosses each other at the same points as the bulk bands at the Fermi level and disappear beyond $U_s$, which indicates appearance of a normal insulator.  }
\label{5}
\end{figure}

\begin{figure}
\centering
\includegraphics[width=8.5cm,scale=1.0]{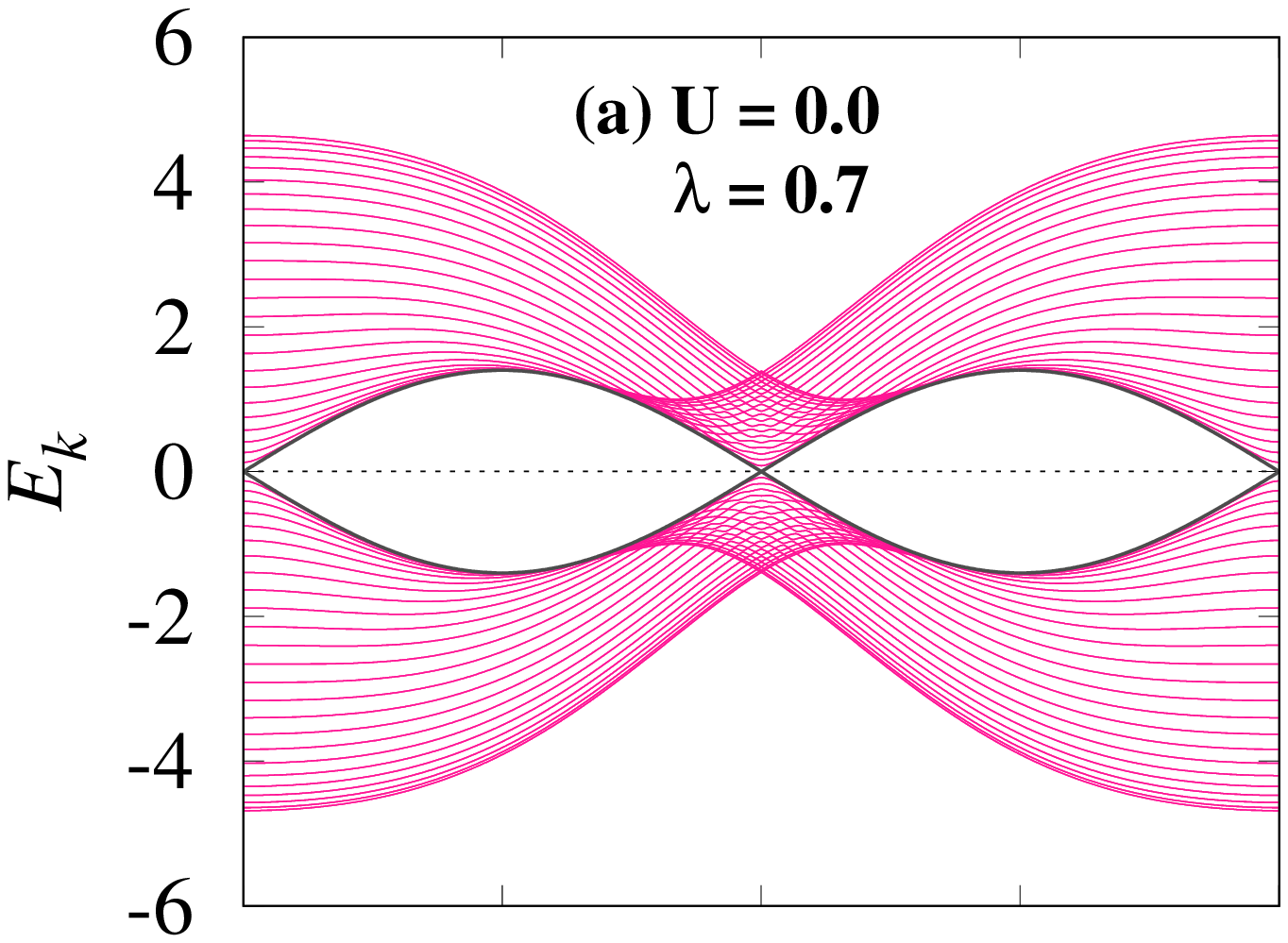}\\
\vspace*{-0.7cm}
\includegraphics[width=8.5cm,scale=1.0]{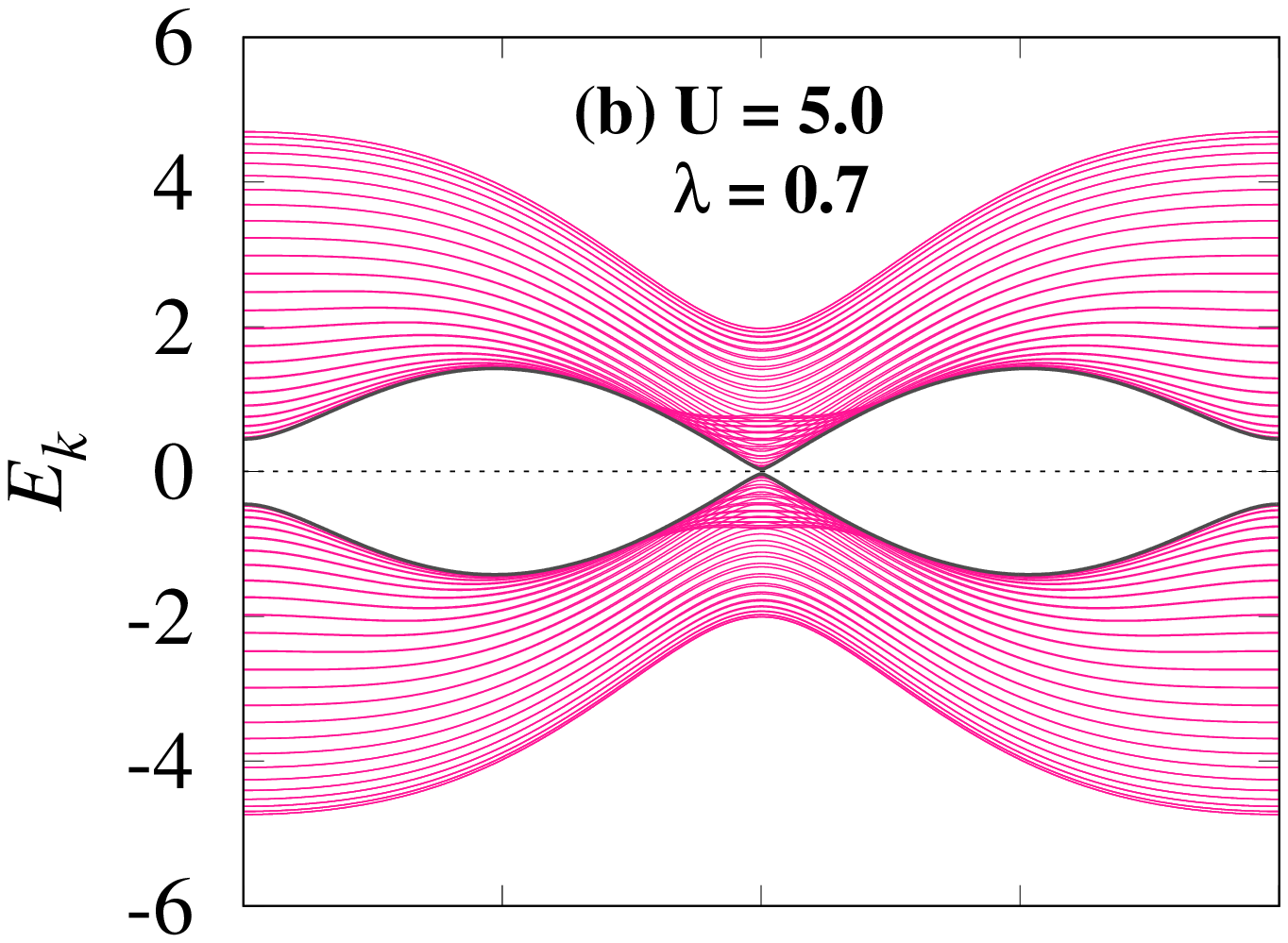} \\
\vspace*{-0.7cm}
\includegraphics[width=8.5cm,scale=1.0]{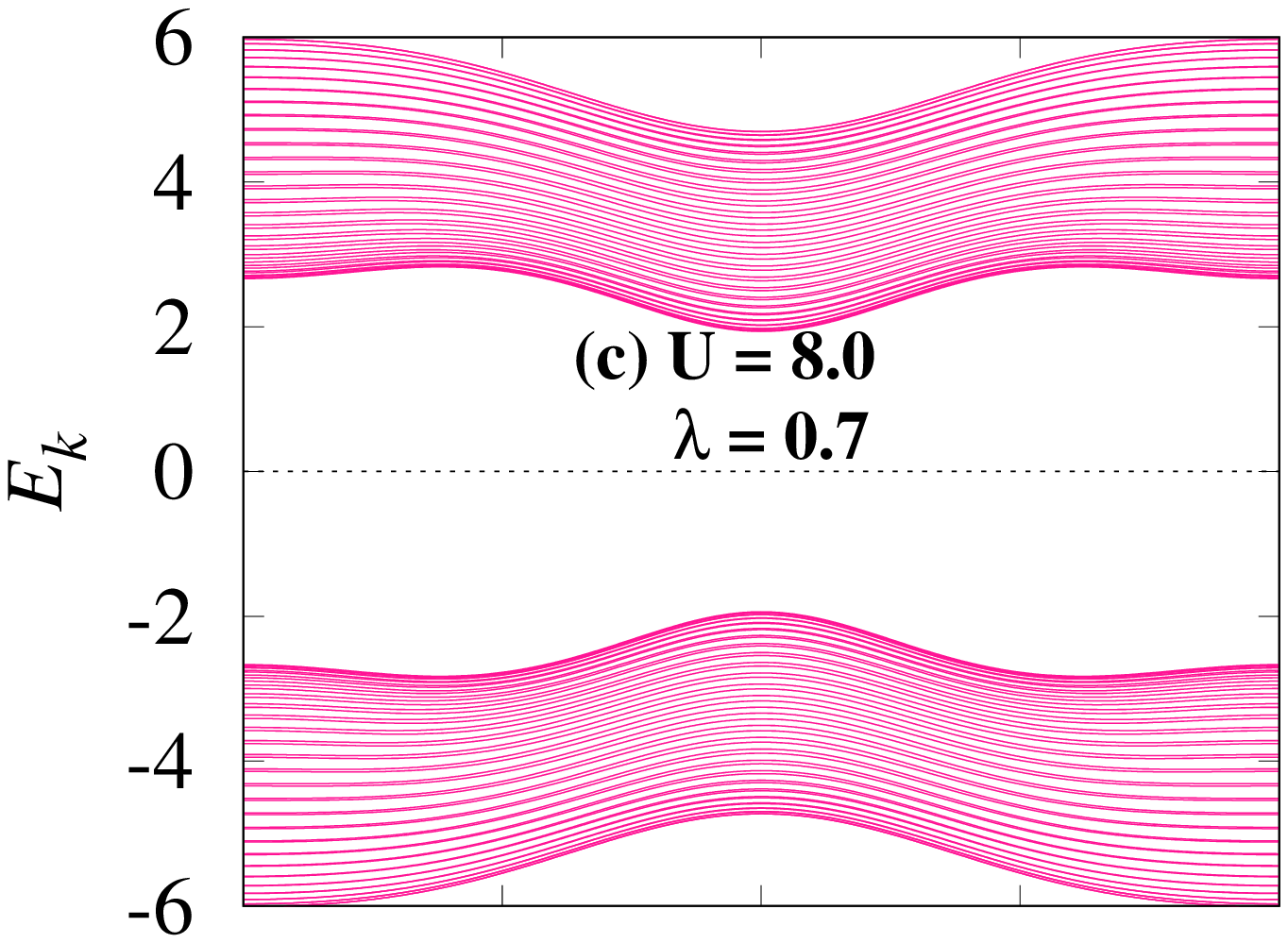} \\
\vspace*{-0.5cm}
\caption{Edge-state dispersions for a ribbon of width $W = 50$ oriented along $x$ direction compared to the bulk bands, which are  projected onto one dimensional Brillouin zone for $\lambda=0.7$ and (a) $U=0.0$, (b) $5.0$ and (c) $8.0$.}
\label{6}
%\end{center}
\end{figure}
Fig.~\ref{5} shows the edge states obtained for three different cases when the number of chains is even. The results are same also for an odd number of chains. In the DSM state without AFM order, when the Hubbard interaction $U = 0$, the edge states cross the Fermi energy at $k_y$ = 0, $\pi$ and $2\pi$. However, in the DSM state with AFM order, the crossing points shift away towards a point in between $k_y$ = 0 and $\pi$. There exists a special $U_s$ for a given value of $\lambda$, where the crossing coincides. Beyond that $U_s$, \textit{i.e.} in the insulating state, there are no states crossing the Fermi energy, indicating the appearance of a normal insulator. 

Next, we consider a ribbon of width $W$ lying along the $x$-direction. The Hamiltonian matrix corresponding to this ribbon also has size $2W \times 2W$ and it is given by
\begin{equation}
    H_{Rbx} ({\bf k}) = 
    \begin{pmatrix}
        H^{\prime}_{1+}  &  H^{\prime}_2  &  H^{\prime}_3  &  O  &  \cdots \\
        H^{\prime \dagger}_2  &  H^{\prime}_{1-}  &  H^{\prime}_2  &  H^{\prime}_3  &  \cdots \\
        H^{\prime \dagger}_{3}  &  H^{\prime \dagger}_2  &  H^{\prime}_{1+}  &  H^{\prime}_2  & \cdots \\
        O  &  H^{\prime \dagger}_3  &  H^{\prime \dagger}_{2}  &  H^{\prime}_{1-}  &  \cdots \\
        \vdots  &  \vdots  &  \vdots  &  \vdots  &  \ddots \\
    \end{pmatrix},
\end{equation}

where 

\begin{equation*}
    H^{\prime}_{1 \pm} = 
    \begin{pmatrix}
       0  &  \pm (-2 i \lambda \sin k_x + \Delta_x)  \\
        \pm (2 i \lambda \sin k_x + \Delta_x)  &  0 \\
    \end{pmatrix},
\end{equation*}
\begin{equation*}
    H^{\prime}_2 = 
    \begin{pmatrix}
        2t \cos (k_x/2)  &  0  \\
        0  &  2t \cos (k_x/2)  \\
    \end{pmatrix}
\end{equation*}
and
\begin{equation*}
    H^{\prime}_3 = 
    \begin{pmatrix}
        0  &  i\lambda  \\
        i\lambda  &  0  \\
    \end{pmatrix}.
\end{equation*}
 $H^{\prime}_{1 \pm}$, $H^{\prime}_{2 }$ and $H^{\prime}_{3 }$ are matrices as described before except that now they are part of Hamiltonian for a ribbon oriented along $x$ axis.

\begin{figure} 
\centering
\includegraphics[width=0.9\linewidth]{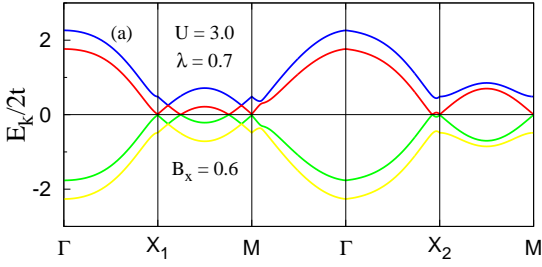}\\
\caption{Electronic dispersion is plotted along the high symmetry directions (a) when the magnetic field is applied along the $x$ axis for the parameters $U=3.0$, $\lambda = 0.7$, and $B_x$ = 0.6. The Dirac points are split into Weyl nodes along X$_1$-M and an additional pair of Weyl nodes emerge along $\Gamma$-X$_2$ as a result of perturbation. }
\label{7}
%\end{center}
\end{figure} 
\begin{figure} 
\centering
\includegraphics[width=0.80\linewidth]{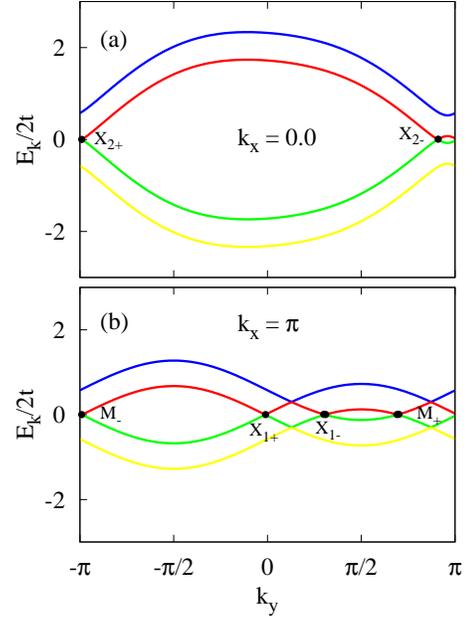}
\vspace{-0.5cm}
\caption{To highlight the Weyl points, the dispersion in Fig~\ref{7} is replotted for (a) $k_x = 0$ and (b) $k_x = \pi$ separately when $k_y$ is varied from $-\pi$ to $\pi$. In addition to the two DPs, which are split into two Weyl points, an additional pair of Weyl points may also be noticed.}
\label{8}
%\end{center}
\end{figure}

Fig.~\ref{6} shows the edge states obtained when the chains as well as the magnetic moments are oriented along the $x$ direction. In the DSM state without any AFM order, the nature of the edge state, as expected, is the same as the case when the ribbon was oriented along the $y$ axis. However, we find that the edge state crosses the Fermi level at $k_x$ = $\pi$ in the DSM state with AFM order. The location of the crossing does not change upon increasing $U$, which is not unusual as the DPs are found along high-symmetry direction for $k_x = \pi$ although the size of the magnetic moments increases with interaction. Finally, the crossing disappears at $U_s$ and beyond.

\begin{figure}
\includegraphics[width=0.9\linewidth]{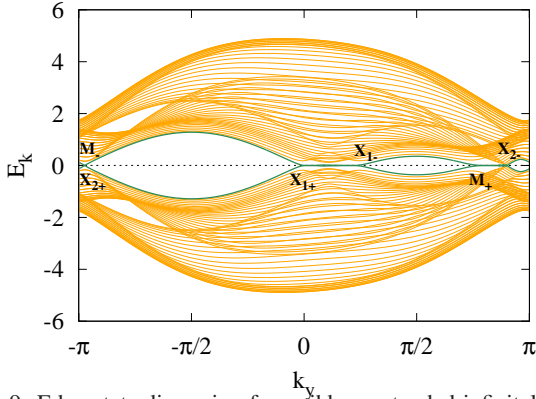}
\centering
\vspace{-0.9cm}
\caption{Edge-state dispersion for a ribbon extended infinitely along $y$ direction and having a finite width $W = 50$ in the presence of magnetic field. The parameters are $U = 3.0$, $\lambda = 0.7$, and $B_x = 0.6$.} 
\label{9}
\end{figure}

Next, we address the question of whether a semimetallic state can be stabilized in the presence of a magnetic field. A magnetic field applied along the positive $x$ axis may support the AFM order with magnetic moments along the $x$ axis. However, the symmetry $\mathcal{P}\mathcal{T}$ effected by $i\sigma_y \mathcal{K} \tau_1$ is preserved, where $\mathcal{K}$ is the complex conjugation operator. The consequence of magnetic field on the electronic dispersion can be obtained by incorporating the term
\begin{equation}
 {\mathcal H}_m = B_x \sigma_x \otimes \tau_0 
\end{equation}
into the Hamiltonian (Eq. (3)), where $B_x$ is the magnetic-field intensity. Then, the electronic dispersion is given by
\begin{eqnarray}
 E_{\bf k} &=& \nonumber\\
 &\pm&
\sqrt{ \pm 2\sqrt{B_x^2(\varepsilon^2_{\bf k}+\Delta^{\prime 2}_{x \bf k})}+ B_x^2 + \varepsilon^2_{\bf k} + \Delta^{\prime 2}_{x \bf k} +\Delta^{\prime 2}_{y\bf k} + \Delta^2_z}. \nonumber\\ 
\end{eqnarray}
It is plotted in Fig.~\ref{7} for self-consistently obtained antiferromagnetically ordered state for $U = 3.0$, $\lambda = 0.7$ and $B_x = 0.6$. Note that the magnetic moments are allowed to choose any direction in the self-consistent scheme and they choose $x$ direction for the above set of parameters. All the three DPs in the presence of any exchange and magnetic field are split into Weyl points. One pair of Weyl points X$_{2\pm}$ are located along $k_x = 0$ at $k_{y0}$ determined by the condition
\begin{equation}
 16 \cos^2 k_y/2 + (\Delta_x-2\lambda \sin k_y)^2 =  B^2_x \\
\end{equation}
provided that the solution exists. On the other hand, two pairs of Weyl points X$_{1\pm}$ and M$_{\pm}$ are found at $k^{}_{y0}$s given by $k_{y0} = \arcsin (\pm B_x + \Delta_x)/2\lambda$ and  $k_{y0} = \pi - \arcsin (\pm B_x + \Delta_x)/2\lambda$ along X$_1$-M for $k_x = \pi$ ((Fig.~\ref{8}(a) and (b)). The Weyl points along $k_x = 0$ and  $\pi$ are accidental as they are not protected by any symmetry, particularly, the nonsymmorphic symmetry $\{C_{2y}|0\frac{1}{2}\}$ is already broken. Whether the DPs are split into Weyl points or the state turns into a normal insulator is determined by the Hubbard on-site interaction $U$. The WSM state is stabilized, when $U$ is small and the magnetic field is comparable to SOC, which forces the magnetic moments to orient along the $x$ direction. Moreover, the accidental degeneracy expected to occur along X$_2$-M in the absence of AFM order is absent~\cite{matneeva}. Formation of Weyl points is accompanied by the edge states, which are shown in Fig.~\ref{9} for a ribbon of infinite length along the $y$ axis and having a finite width $W = 50$ along $x$ axis. The edge states are the projections of the Weyl points which are accompanied by the flat bands connecting the Weyl nodes with opposite Chern numbers. They are distinctly seen to be present for each pair of Weyl points.

In order to compute the Chern number, the Hamiltonian in the presence of magnetic field can reduced to a simple form by the following unitary transformation 
\begin{equation}
    U = \frac{1}{\sqrt{2}}
    \begin{pmatrix}
        \sigma_{0}   &   \sigma_{x} \\
        \sigma_{z}   &   -i \sigma_{y}
    \end{pmatrix}.
\end{equation}
After the transformation, the Hamiltonian becomes
\begin{equation}
    \mathcal{H}^{'} = U^{-1} \mathcal{H} U = 
    \begin{pmatrix}
        H_{-}   &    B_x \sigma_0  \\
         B_x  \sigma_0    &    H_{+} 
    \end{pmatrix}.
\end{equation}
Here, $H_{+}=H_{-}^{\dagger}={\bf d \cdot \sigma}$ with ${\bf d}=\{\Delta_{x {\bf k}}^{'},\Delta_{y {\bf k}}^{'},\epsilon_{\bf k}\}$ and $\sigma_i$s are the Pauli matrices. The above Hamiltonian $\mathcal{H}^{'}$ reduces to the block diagonal form in the absence of any external magnetic field. This form of the Hamiltonian can be used to calculate the Chern number for the various Weyl points.
\begin{figure}[hb]
    \centering
    \includegraphics[width=1.0\linewidth]{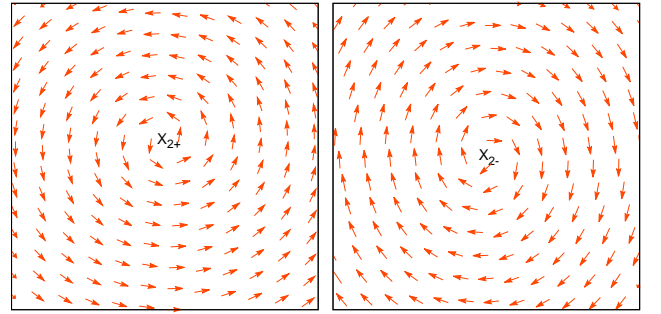}
    \hspace{-0.8cm}
    \caption{Berry connection plotted for a pair of Weyl points $X_{2+}$ and $X_{2-}$ obtained when an external magnetic field is applied along $x$-direction. The Weyl points for the given pair possess winding numbers -1 and +1 depending on the clockwise and anticlockwise rotations.}
    \label{10}
\end{figure}
The Berry connection for a pair of Weyl points $X_{2+}$ and $X_{2-}$ is shown in Fig~\ref{10}. Each Weyl point has a winding number as -1 or +1 depending on the clockwise or counterclockwise rotations of the Berry connection. As expected, the total rotation of a single pair of Weyl point vanishes.

Here, we focused on the role of interaction in stabilizing the magnetically ordered topological semimetallic state within the one-band Hubbard model at half-filling. In particular, we showed that such a state is realized when an in-plane magnetic field is applied along (1, 0) or (0, 1) direction. The consequence of band fillings other than half, the presence of more than one orbital, and nearest-neighbor spin-orbit coupling are issues of significant interest which we discuss below. 

\section{Discussion}
In the current work, the second-neighbor spin-orbit coupling was considered instead of the first-neighbor~\cite{young}. As originally proposed, the topologically-protected DSM state may exist in a non-magnetic state when the second-neighbor spin-orbit coupling and the nonsymmorphic symmetry are present. It may be noted that the direction of magnetic moments depends on the nature of spin-orbit coupling. The self-consistently obtained  magnetic moments are oriented in plane when only the first-neighbor spin-orbit coupling is considered in contrast to their out of plane alignment when only the second-neighbor spin-orbit coupling is considered. It may also be noted that a gap is opened at the Fermi level for half-filling in the case of the former.

There is no significant qualitative change in the phase diagram when the temperature is varied. This is because, in the static meanfield theoretic approach, a phase transition from magnetic to non-magnetic order occurs through the melting of magnetic moment, \textit{i.e.}, the magnetic moment for a given set of parameter decreases with increase in temperature and vanishes at the critical temperature. Therefore, the effect of increase in temperature is expected to shift the phase boundaries, \textit{i.e.} both the phase boundaries, separating DSM-NM and DSM-AFM, and separating DSM-AFM and AFM-I will shift downwards in the $\lambda$ vs $U$ phase diagram. Secondly, for large $U$, a more refined treatment for the phase diagram can be obtained with Gutzwiller approximation (GA). The main consequence of GA is to renormalize $U$, which leads to the reduction in the range of $U$ for which the magnetically ordered phase will be stabilized~\cite{markiewicz}. The effect of renormalization may get enhanced when the frustration in the hopping process is incorporated. A similar effect is expected to take place in the presence of second-neighbor SOC.

A magnetically ordered DSM state within the Hubbard model exists only when two of the four bands ($n = 2$) are completely filled and the spin arrangement is of checkerboard type. Not much is known about a magnetically ordered DSM state for other band fillings. A recent work suggests that a DSM state with stripe-like magnetic order and integer band fillings other than $n = 2$ can also be realized when the band filling $n = 6$, i. e., six out of eight bands are completely filled. While $n = 2$ corresponds to one electron per site in the checkerboard case, $n = 6$ for stripe order corresponds to an average electronic occupancy of three-quarter for each site~\cite{young2}. It will be interesting to see whether the DSM state with striped magnetic order in the presence of both the first- and second-neighbor spin-orbit coupling is stable within the self-consistent mean-field theory.  

The DSM state in the multi-band correlated electron system is of strong current interest and the existence of DPs have been reported in a variety of systems though often far from the Fermi surface. A large volume of studies have focused on the electronic-band structure and system without any magnetic order. However, one widely studied multi-orbital systems with moderate correlations is iron pnictide, which exhibits a striped spin-density wave (SDW) order. The DPs in the SDW state of iron pnictides are not far away from the Fermi level and they are believed to play an important role in the transport properties. Although the symmetries which protect the DPs are different. In particular, the stability of the Dirac cones and nodes in the SDW state without any nonsymmorphic symmetry is a consequence of three symmetries, (i) collinear arrangement of magnetic moments, (ii) inversion symmetry about an iron atom, and combined time reversal and inversion of magnetic moments~\cite{wan}. Since the the DPs are not far away from the Fermi surface, there is a possibility of tuning the location of these DPs by parameters such as orbital-splitting by applying mechanical pressure~\cite{garima}.

\section{Conclusion}
To conclude, we have examined the possible  existence of Dirac semimetallic state with AFM ordering within one-orbital Hubbard model with second-neighbor spin-orbit coupling and nearest-neighbor hopping. In the non-symmorphic symmetry, when the magnetic moments are directed along the $x$ axis in the antiferromagnetic arrangement, we obtain the phase diagram in the interaction vs spin-orbit coupling parameter space. The nature of edge states is uncovered for different relative orientations of magnetic moments with respect to ribbon geometry. Our findings also suggest that the topological semimetallic state with AFM order may not be stabilized unless the magnetic moments are forced to align along the line segments joining nearest neighbor atoms in a particular sublattice. On the other hand, Weyl semimetal is stabilized when magnetic field is applied along a direction same as that of magnetic moments. 

\section*{ACKNOWLEDGEMENTS} 
D.K.S. was supported through DST/NSM/R\&D\_HPC\_Applications/2021/14 funded by DST-NSM and start-up research grant SRG/2020/002144 funded by DST-SERB.

\end{document}